\documentclass[12pt]{article}
\usepackage{amsfonts,amssymb,epsfig,amsmath,mathtools}
\usepackage{color}

\renewcommand{\baselinestretch}{1.2}

\newcommand{\be}{\begin{eqnarray}}
\newcommand{\ee}{\end{eqnarray}}

\newcommand{\bn}{\begin{enumerate}}
\newcommand{\en}{\end{enumerate}}

\begin{document}

\makeatletter \@addtoreset{equation}{section} \makeatother
\renewcommand{\theequation}{\thesection.\arabic{equation}}
\renewcommand{\thefootnote}{\alph{footnote}}

\begin{titlepage}

\begin{center}

\hfill {\tt SNUTP18-006}\\
\hfill{\tt KIAS-P18076}\\

\vspace{2cm}

{\Large\bf Entropy functions of BPS black holes in AdS$_4$ and AdS$_6$}

\vspace{2cm}

\renewcommand{\thefootnote}{\alph{footnote}}

{\large Sunjin Choi$^1$, Chiung Hwang$^{2,3}$, Seok Kim$^1$ and June Nahmgoong$^1$}

\vspace{0.7cm}

\textit{$^1$Department of Physics and Astronomy \& Center for
Theoretical Physics,\\
Seoul National University, Seoul 08826, Korea.}\\

\vspace{0.2cm}

\textit{$^2$School of Physics, Korea Institute for Advanced Study,
Seoul 02455, Korea.}\\

\vspace{0.2cm}

\textit{$^3$Dipartimento di Fisica, Universit\`{a} di Milano-Bicocca \& INFN, \\
Sezione di Milano-Bicocca, I-20126 Milano, Italy.}\\

\vspace{0.7cm}

E-mails: {\tt csj37100@snu.ac.kr, chwang@kias.re.kr, \\
skim@phya.snu.ac.kr, earendil25@snu.ac.kr}

\end{center}

\vspace{1cm}

\begin{abstract}

We find the entropy functions of supersymmetric black holes in AdS$_4$ and AdS$_6$
with electric charges and angular momenta. Extremizing these functions, one obtains the
entropies and the chemical potentials of known analytic black hole solutions.

\end{abstract}

\end{titlepage}

\renewcommand{\thefootnote}{\arabic{footnote}}

\setcounter{footnote}{0}

\renewcommand{\baselinestretch}{1}

\tableofcontents

\renewcommand{\baselinestretch}{1.2}

\section{Introduction}

Understanding black holes \cite{Hawking:1982dh,Witten:1998zw,Aharony:2003sx}
is an important subject in AdS/CFT \cite{Maldacena:1997re}. In models with supersymmetry,
one expects that quantitative analysis at strong coupling would be easier in the BPS
sectors of SCFTs. Supersymmetric AdS black holes correspond to thermal ensembles
of BPS states, carrying angular momenta and also internal charges (electric
charges in AdS). In AdS$_d$ with $d>3$, supersymmetric black holes have very
complicated structures. First of all, it is known that there are no BPS black holes with
electric charges only, at zero angular momenta. This is because in the dual field theory,
the local BPS operators will reduce to chiral rings which do not have enough numbers
of microstates to form black holes: e.g. see \cite{Kinney:2005ej} for the case with $d=5$.
With nonzero angular momenta, the solutions appear very involved. See, e.g.
\cite{Kostelecky:1995ei,Cvetic:2005zi} for $d=4$,
\cite{Gutowski:2004ez,Gutowski:2004yv,Chong:2005da,Kunduri:2006ek} for $d=5$,
\cite{Chow:2008ip} for $d=6$, and \cite{Chong:2004dy,Chow:2007ts} for $d=7$.

In \cite{Hosseini:2017mds,Hosseini:2018dob}, simple underlying structures of
BPS black holes were discovered in AdS$_5$ and AdS$_7$.
Firstly, one can obtain the entropies and the chemical potentials of known BPS black
holes \cite{Gutowski:2004ez,Gutowski:2004yv,Chong:2005da,Kunduri:2006ek}
in AdS$_5\times S^5$ by extremizing the following entropy
function \cite{Hosseini:2017mds},
\begin{equation}\label{AdS5}
  S(\Delta_I,\omega_i)=\frac{N^2\Delta_1\Delta_2\Delta_3}{2\omega_1\omega_2}
  +\sum_{I=1}^3\Delta_I Q_I+\sum_{i=1}^2\omega_i J_i\ ,
\end{equation}
subject to the constraint $\Delta_1+\Delta_2+\Delta_3-\omega_1-\omega_2=2\pi i$.
Here $Q_I$ and $J_i$ are $U(1)^3\subset SO(6)$ electric charges and
$U(1)^2\subset SO(4)$ angular momenta, respectively. The Bekenstein-Hawking
entropy of the black hole is the extremal value of ${\rm Re}(S)$, at one of the extremum
solutions for $\Delta_I,\omega_i$ \cite{Hosseini:2017mds}. The black hole
chemical potentials are the extremal values of ${\rm Re}(\Delta_I)$, ${\rm Re}(\omega_i)$
\cite{Choi:2018hmj}. Similarly, the properties of known BPS black holes \cite{Chong:2004dy,Chow:2007ts} in AdS$_7\times S^4$ can be understood by extremizing
the following entropy function \cite{Hosseini:2018dob},
\begin{equation}\label{AdS7}
  S(\Delta_I,\omega_i)=-\frac{N^3(\Delta_1\Delta_2)^2}{24\omega_1\omega_2\omega_3}
  +\sum_{I=1}^2\Delta_I Q_I+\sum_{i=1}^3\omega_i J_i\ ,
\end{equation}
subject to the constraint $\Delta_1+\Delta_2-\omega_1-\omega_2-\omega_3=2\pi i$.
$Q_I$ and $J_i$ are $U(1)^2\subset SO(5)$ electric charges and $U(1)^3\subset SO(6)$
angular momenta. These extremely simple formulae encode apparently complicated properties
of supersymmetric AdS black holes. They triggered interesting follow-up works aiming at
microscopic accounts for these black holes \cite{Cabo-Bizet:2018ehj,Choi:2018hmj,Choi:2018vbz}.
In particular, \cite{Choi:2018hmj} derived (\ref{AdS5}) for large AdS$_5$ black holes,
from the index of 4d $\mathcal{N}=4$ Yang-Mills theory. (\cite{Choi:2018hmj} also
found a generalization of \cite{Hosseini:2017mds}
in a different charge sector.) \cite{Choi:2018hmj,Nahmgoong:2019hko} also provided an anomaly-based
microscopic discussion which leads to (\ref{AdS7}) for large AdS$_7$ black holes.
Therefore, the simple functions (\ref{AdS5}) and (\ref{AdS7}) provided very useful
inspirations for microscopic studies.

We thus find that it will be very helpful to have entropy function formalisms
for supersymmetric AdS black holes in other dimensions. In this note, we provide
such functions in AdS$_4$ and AdS$_6$, simplifying the apparently complicated
structures of known black hole solutions. These led to better microscopic
understandings based on CFT$_3$ \cite{Choi:2019zpz} and CFT$_5$ duals \cite{Choi:2019miv}. In particular, \cite{Choi:2019zpz} and \cite{Choi:2019miv} statistically accounted for large BPS black holes in AdS$_4$ and AdS$_6$, from the indices of CFT$_3$ and CFT$_5$ duals.

The rest of this note is organized as follows. Section 2 summarizes the properties
of known supersymmetric black holes in AdS$_4\times S^7$, and show that an entropy function we
suggest encodes these properties. Section 3 makes similar studies with supersymmetric
AdS$_6$ black holes. Section 4 concludes with remarks.

\section{AdS$_4$ black holes\label{AdS4}}

\subsection{Black hole solutions\label{AdS4-BH}}

We study the supersymmetric black holes in AdS$_4\times S^7$ of \cite{Cvetic:2005zi}.
These are obtained by taking supersymmetric limits of \cite{Chong:2004na}, also demanding
the existence of smooth horizons.

Black holes in AdS$_4\times S^7$ can carry six kinds of conserved
quantities: mass (or energy) $E$, angular momentum $J$ on $S^2$ of global AdS$_4$, and four
Cartan charges $Q_I$ ($I=1,2,3,4$) of $SO(8)$ symmetry on $S^7$. The last four conserved
quantities $Q_I$ appear in 4d gravity as $U(1)^4$ electric charges. The convention of \cite{Cvetic:2005zi} for $Q_I$ is to take four angular momenta acting on the orthogonal 2-planes of $\mathbb{R}^8$ related to $S^7$. The most general black holes known to date have pairwise
equal electric charges, $Q_1=Q_3$, $Q_2=Q_4$.
With the last charge restrictions, the four conserved quantities $E,J,Q_1,Q_2$
are labeled by four parameters $m,a,\delta_1,\delta_2$ as \cite{Cvetic:2005zi}
\begin{eqnarray}\label{AdS4-charge}
  &&E=\frac{m}{2G\Xi^2}(\cosh 2\delta_1+\cosh 2\delta_2)\ \ ,\ \ \
  J=\frac{ma}{2G\Xi^2}(\cosh 2\delta_1+\cosh 2\delta_2)\ \ ,\nonumber\\
  &&Q_1=Q_3=\frac{m}{4G\Xi}\sinh 2\delta_1\ \ ,\ \ \
  Q_2=Q_4=\frac{m}{4G\Xi}\sinh 2\delta_2\ ,
\end{eqnarray}
where $\Xi=1-a^2g^2$. The entropy is given by
\begin{equation}
  S=\frac{\pi (r_1r_2+a^2)}{G\Xi}\ ,
\end{equation}
where $r_i=r_++2m\sinh^2\delta_i$. $r=r_+$ is the location of the event horizon.
$G$ is the 4d Newton constant, which will be replaced by microscopic parameters later. (In \cite{Cvetic:2005zi}, all charges and entropy are computed omitting the overall $\frac{1}{G}$ factor, or at $G=1$. E.g. the entropy is computed by dividing the horizon area by $4$, rather than $S=\frac{A}{4G}$.) $g$ is a parameter of the 4d gauged supergravity, and is related to the radius $\ell$ of AdS$_4$
as $g=\ell^{-1}$.

The BPS limit of these black holes is given by
\begin{equation}\label{BPS-AdS4}
  e^{2\delta_1+2\delta_2}=1+\frac{2}{ag}\ ,
\end{equation}
which corrects a typo of \cite{Cvetic:2005zi}. Only after this correction,
the BPS relation
\begin{equation}\label{BPS-energy-AdS4}
  E=gJ+\sum_{I=1}^4Q_I=gJ+2Q_1+2Q_2
\end{equation}
is met. A further condition to have a regular horizon is $\Delta_r=0$ having a double root at $r=r_+$. (See \cite{Cvetic:2005zi} for the definition of the function $\Delta_r$.)
This happens only after a further tuning of $m$. After the tuning,
the horizon location $r=r_+$ is given by
\begin{equation}\label{horizon-location}
  r_+=\frac{2m\sinh\delta_1\sinh\delta_2}{\cosh(\delta_1+\delta_2)}\ ,
\end{equation}
when $m$ satisfies
\begin{equation}\label{regular-AdS4}
  (mg)^2=\frac{\cosh^2(\delta_1+\delta_2)}
  {e^{\delta_1+\delta_2}\sinh^3(\delta_1+\delta_2)\sinh(2\delta_1)\sinh(2\delta_2)}\ .
\end{equation}
This again corrects the formula
$mg=\frac{\cosh(\delta_1+\delta_2)}{e^{\frac{\delta_1+\delta_2}{2}}
\sinh^2(\delta_1+\delta_2)\sinh(2\delta_1)\sinh(2\delta_2)}$ of \cite{Cvetic:2005zi}. 
The typos found in this paragraph are also reported in \cite{Chow:2013gba}.

Taking the BPS limit, the entropy of the supersymmetric black hole is given by
\begin{equation}\label{BPS-AdS4-entropy}
S=\frac{2\pi}{g^2G(e^{2\delta_1+2\delta_2}-3)}\ .
\end{equation}
The two conditions (\ref{BPS-AdS4}), (\ref{regular-AdS4}) leave two independent parameters among $m,a,\delta_1,\delta_2$. Even after restricting $E$ as (\ref{BPS-energy-AdS4}) due to the BPS condition, the remaining charges $Q_1,Q_2,J$ satisfy a relation. Together with
$S$, we find the following two relations after taking the BPS limit:
\begin{equation}
\begin{aligned}
&\left(\frac{2Q_1}{g}+\frac{2Q_2}{g}\right)S=\frac{\pi}{g^2G}J, \\
&S^2+\frac{\pi}{g^2G}S-4\pi^2 \frac{2Q_1}{g}\frac{2Q_2}{g}=0.
\end{aligned}
\end{equation}
Since these equations determine $S$ twice, one will get a charge relation between $Q_1, Q_2, J$ from the compatibility of two equations. Explicitly, we insert the solution of the first equation to the second equation, demanding two equations have the same solution for $S$. Then, taking the unique positive solution assuming $Q_1,Q_2,J>0$, one obtains
\begin{equation}\label{AdS4-bh-cr}
\begin{aligned}
&S= \frac{\pi}{g^2G} \frac{J}{\left(\frac{2Q_1}{g}+\frac{2Q_2}{g}\right)}\ , \\
&J=\frac{1}{2} \left(\frac{2 Q_1}{g} +\frac{2 Q_2}{g} \right) \left(-1+\sqrt{1+16g^4G^2\frac{2Q_1}{g} \frac{2Q_2}{g}}  \right)\ .
\end{aligned}
\end{equation}
Thus, we have explicitly found the charge relation between $Q_1, Q_2, J$.

The black hole chemical potentials and the free energy $F$ satisfy
\begin{equation}
  S=-T^{-1}F(T)+T^{-1}E-T^{-1}\Omega J-T^{-1}\sum_{I=1}^4\Phi_I Q_I\ ,
\end{equation}
where $T$ is the temperature,  $\Omega$ is the angular velocity, and $\Phi_i$'s are the electrostatic potentials. The chemical potentials are evaluated on the horizon. In the BPS limit we are interested in,
\begin{equation}
  T=\frac{\Delta_r^\prime}{4\pi(r_1r_2+a^2)}\rightarrow 0
\end{equation}
because $\Delta_r$ has a double root at the horizon. On the other hand, as one inserts the value of the variables in the BPS limit,
$a=\frac{2}{g(e^{2\delta_1+2\delta_2}-1)}$, $mg$ given by (\ref{regular-AdS4}), and then the horizon location $r\rightarrow r_+$ (\ref{horizon-location}), one finds
\begin{equation}
  \Omega=\frac{a(1+g^2r_1r_2)}{r_1r_2+a^2}\rightarrow g\ ,\ \
  \Phi_1=\frac{mr_2 \sinh (2\delta_1)}{r_1r_2+a^2}\rightarrow 1\ ,\ \
  \Phi_2=\frac{mr_1 \sinh (2\delta_2)}{r_1r_2+a^2}\rightarrow 1\ .
\end{equation}
Defining $\Delta E$ by $E=\Delta E+2Q_1+2Q_2+gJ$, one finds that
\begin{equation}
  S=-T^{-1}F(T)+T^{-1}\Delta E-T^{-1}(\Omega-g)J-T^{-1}\sum_{I=1}^4(\Phi_I-1)Q_I\ .
\end{equation}
The BPS limit satisfies $T\rightarrow 0$, $\Delta E\rightarrow 0$.
One first finds that
\begin{equation}
  \omega=-\lim_{T\rightarrow 0}\left(T^{-1}(\Omega-g)\right)
  \ ,\ \ \Delta_I=-\lim_{T\rightarrow 0}\left(T^{-1}(\Phi_I-1)\right)
\end{equation}
are well defined in the BPS limit, by explicitly computing them (although the expressions
are very complicated). Since $S$ is also finite in this limit,
the `BPS free energy' $F_{\textrm{BPS}}\equiv\lim_{T\rightarrow 0}(T^{-1}(F-\Delta E))$
should also be well defined. So one finds
\begin{equation}
  S=-F_{\textrm{BPS}}+\omega J+\sum_{I=1}^4\Delta_IQ_I
\end{equation}
in the BPS limit. $-F_{\rm BPS}$ is to be interpreted as $\log Z$, where $Z$ is
the BPS partition function of this system.
We again stress that the BPS limit is taken by first inserting $ag\rightarrow\frac{2}{e^{2\delta_1+2\delta_2}-1}$, $mg\rightarrow\sqrt{\frac{(\coth(\delta_1+\delta_2)-1)
\coth^2(\delta_1+\delta_2)}{\sinh(2\delta_1)\sinh(2\delta_2)}}$ and then $r\rightarrow\frac{2m\sinh\delta_1 \sinh\delta_2}{\cosh(\delta_1+\delta_2)}$. This
results in quite complicated expressions for $\omega,\Delta_i$. After taking the
BPS limit, one can show that they satisfy
\begin{equation}\label{ch-rel-AdS4}
  \Delta_1+\Delta_2=\frac{1}{g}\omega \quad \Rightarrow \quad \sum_{I=1}^4 \frac{g}{2}\Delta_I-\omega=0\ .
\end{equation}
This is an alternative statement of the charge relation between $Q_1,Q_2,J$.

\subsection{Entropy function\label{AdS4-entropy}}

We now present an entropy function, whose suitable Legendre transformation in
$\Delta_I,\omega$ yields the entropy $S(Q_I,J)$ and the BPS chemical potentials
of the supersymmetric black holes. Our entropy function $S(\Delta_I,\omega;Q_I,J)$
is given by
\begin{equation}\label{entropy-ftn-AdS4}
  S(\Delta_I,\omega;Q_I,J)
  =- i\frac{4\sqrt{2}N^{\frac{3}{2}}}{3}\frac{\sqrt{\Delta_1 \Delta_2 \Delta_3 \Delta_4}}{\omega}
  +\omega J+\sum_{I=1}^4\Delta_IQ_I\ .
\end{equation}
We extremize $S$ in $\Delta_I,\omega$ with the constraint
\begin{equation}\label{constraint-AdS4}
  \Delta_1+\Delta_2+\Delta_3+\Delta_4-\omega=2\pi i\ .
\end{equation}
A microscopic derivation of the entropy function (\ref{entropy-ftn-AdS4}) from the CFT$_3$ dual was studied in \cite{Choi:2019zpz}, in the Cardy limit $\omega \to 0$. Just like AdS$_5$, AdS$_7$ black holes analyzed in \cite{Choi:2018hmj}, the
constraint (\ref{constraint-AdS4}) is given an interpretation in \cite{Choi:2019zpz}.
Here the number of M2-branes $N$ is related to the 4d Newton constant $G$ as follows:
\begin{equation}
G_{11}=16\pi^7 {\ell_\textrm{P}}^9, \; \ell_{S^7}=2\ell=
\ell_\textrm{P}(2^5\pi^2 N)^{1/6} \; \Rightarrow \; \frac{1}{g^2G}=\frac{\textrm{vol}(S^7)}{g^2G_{11}}=\frac{2\sqrt{2}}{3} \frac{N^{3/2}}{g^2\ell^2}=\frac{2\sqrt{2}}{3} N^{3/2}\ .
\end{equation}
$\ell_{\rm P}$ is the 11d Planck scale, $\ell_{S^7}$ is the radius of $S^7$, and
$\ell$ is the AdS$_4$ radius as defined in the previous subsection.
We claim that the resulting extremal value of ${\rm Re}(S)$ is the entropy of supersymmetric
black holes. We shall check this against the known solutions summarized in the previous
subsection, at $Q_1=Q_3$, $Q_2=Q_4$ (which is equivalent to $\Delta_1=\Delta_3$, $\Delta_2=\Delta_4$). Here, note that the chemical potentials
$\Delta_I$, $\omega$ are all complexified. With complex $\Delta_I$, the square root
$\sqrt{\Delta_1\Delta_2\Delta_3\Delta_4}$ in (\ref{entropy-ftn-AdS4}) should be understood as to take the argument of $\Delta_1 \Delta_2 \Delta_3 \Delta_4$ in the principal branch $(-\pi, \pi)$ \cite{Choi:2019zpz}.

We show our claim by extremizing $S$, subject to the constraint (\ref{constraint-AdS4}). We introduce the Lagrange multiplier $\lambda$ and extremize
\begin{equation}\label{S-AdS4}
  S=- i\frac{4\sqrt{2}N^{\frac{3}{2}}}{3}\frac{\sqrt{\Delta_1\Delta_2\Delta_3\Delta_4}}{\omega}
  +\omega J+\sum_{I=1}^4\Delta_IQ_I+\lambda\left(\sum_{I=1}^4 \Delta_I-\omega-2\pi i\right)\ .
\end{equation}
The extremum conditions are given by
\begin{eqnarray}\label{extremal-AdS4}
  \lambda+Q_I&=& i\frac{4\sqrt{2}N^{\frac{3}{2}}}{3\omega}
  \frac{\sqrt{\Delta_1\Delta_2\Delta_3\Delta_4}}{2\Delta_I}\ \ \  (I=1,\cdots,4)\ , \\
  \lambda-J&=& i\frac{4\sqrt{2}N^{\frac{3}{2}}}{3\omega^2}\sqrt{\Delta_1\Delta_2\Delta_3\Delta_4}\ .\nonumber
\end{eqnarray}
Inserting these charges into (\ref{S-AdS4}), to eliminate the appearances of $Q_I,J$,
one obtains
\begin{equation}
S=-2\pi i \lambda\ .
\end{equation}
Multiplying the four equations on the first line of \eqref{extremal-AdS4}, one finds
\begin{equation}
(\lambda+Q_1)(\lambda+Q_2)(\lambda+Q_3)(\lambda+Q_4)= \frac{64 N^6}{81\omega^4} \Delta_1\Delta_2\Delta_3\Delta_4 =- \frac{2N^3}{9} (\lambda-J)^2\ .
\end{equation}
So one obtains a very useful expression,
\begin{equation}\label{S-polynomial}
\left(\frac{S}{2\pi i}-Q_1\right)\left(\frac{S}{2\pi i}-Q_2\right)\left(\frac{S}{2\pi i}-Q_3\right)\left(\frac{S}{2\pi i}-Q_4\right)=- \frac{2N^3}{9} \left(\frac{S}{2\pi i}+J\right)^2\ .
\end{equation}
One needs care to treat the above expression. While the above is the quartic equation in $S$, only the half of them are the true solutions to \eqref{extremal-AdS4} satisfying the constraint (\ref{constraint-AdS4}). The other halves are the extraneous solutions. Hence, after solving the above equation, one should check whether the resulting solution is a true one.

After extremizing the entropy function, one would generally obtain complex solutions for
$S$ by solving (\ref{S-polynomial}). Along the spirit of \cite{Choi:2018hmj}, we shall generally
regard ${\rm Re}(S)$ as the entropy at the extremum. See \cite{Choi:2018hmj,Choi:2019zpz} for the
interpretation of the imaginary part. However, we are primarily interested in comparing
our results against the known black hole solutions of section \ref{AdS4-BH}. Therefore, we impose
the charge relation of these black holes and compare the thermodynamic quantities on that
surface only. Somewhat remarkably, the charge relation of known black holes will turn out
to be ${\rm Im}(S)=0$ at the extremum of our entropy function. So from now on, we demand
the existence of a real solution for $S$ in (\ref{S-polynomial}), and compare the results
with the known black holes.
Demanding real $S$ for real charges $Q_1, Q_2, Q_3, Q_4, J$,
the complex equation (\ref{S-polynomial}) is separated into two real equations as follows:
\begin{eqnarray}
  \frac{1}{16\pi^4}S^4 - \frac{\sum_{I < J} Q_I Q_J}{4\pi^2} S^2 +Q_1Q_2Q_3Q_4
  &=& \frac{N^3}{18\pi^2} S^2-\frac{2N^3J^2}{9}\ , \nonumber\\
  -\frac{\sum_I Q_I}{8\pi^3}S^3 + \frac{\sum_{I<J<K}Q_IQ_JQ_K}{2\pi} S
  &=& \frac{2N^3J}{9\pi} S\ .
\end{eqnarray}
These equations determine $S$ twice as functions of charges. From the compatibility of two equations, one will get a relation of $Q_I, J$. Explicitly, one may take the unique positive solution of the second equation and insert it to the first equation, to obtain the charge relation. One can check that this solution is a true solution satisfying \eqref{extremal-AdS4} and (\ref{constraint-AdS4}).

To compare with known black holes summarized in section \ref{AdS4-BH},
we set $Q_1=Q_3$, $Q_2=Q_4$. Then, taking the unique positive solution
assuming $Q_1,Q_2,J>0$, one obtains
\begin{eqnarray}
  S&=& \frac{2\pi}{3} \sqrt{\frac{9Q_1Q_2(Q_1+Q_2)-2N^3 J}{Q_1+Q_2}},\nonumber \\
  0&=&2N^3J^2+2N^3(Q_1+Q_2)J-9Q_1Q_2(Q_1+Q_2)^2\ .
\end{eqnarray}
These can be rearranged as
\begin{eqnarray}
  S&=&\frac{2\sqrt{2}\pi N^{\frac{3}{2}}}{3} \frac{J}{Q_1+Q_2}
  = \frac{\pi}{g^2G}\frac{J}{Q_1+Q_2}\ , \\
  J&=&\frac{1}{2} (Q_1+Q_2) \left(-1 + \sqrt{1+\frac{18}{N^3}Q_1Q_2}\right)
  = \frac{1}{2} (Q_1+Q_2) \left(-1 + \sqrt{1+16g^4G^2Q_1Q_2}\right)\ .\nonumber
\end{eqnarray}
One can easily check that this solution indeed satisfies \eqref{extremal-AdS4} and (\ref{constraint-AdS4}), i.e. it is not an extraneous solution.
The above expressions are exactly the same as \eqref{AdS4-bh-cr}, which we obtained from the supersymmetric black holes.
Note that the charges and chemical potentials of the entropy function \eqref{S-AdS4} are related to those of supersymmetric black holes as
\begin{equation}
\begin{aligned}
&S_{\textrm{BH}}=S, \; J_{\textrm{BH}}=J, \; \frac{2}{g}Q_{I, \textrm{BH}}=Q_{I}\ ,\\
&\omega_{\textrm{BH}}=\textrm{Re}(\omega), \; \frac{g}{2}\Delta_{I, \textrm{BH}}=\textrm{Re}(\Delta_{I}).
\end{aligned}
\end{equation}
Here, the subscripts `BH' denote the black hole quantities, while the others are the quantities used in the entropy function. The second line can be shown by a rather straightforward but
tedious calculus. One also finds that the relation between the chemical potentials in the entropy function \eqref{constraint-AdS4} is equivalent to that of the supersymmetric black holes \eqref{ch-rel-AdS4}.

To summarize, our entropy function \eqref{S-AdS4} indeed reproduces the Bekenstein-Hawking entropy of the supersymmetric AdS$_4$ black holes \eqref{BPS-AdS4-entropy} and the corresponding charge/chemical potential relations \eqref{AdS4-bh-cr}, \eqref{ch-rel-AdS4}, at
$Q_1=Q_3$, $Q_2=Q_4$ where solutions are known. 
Recently, $4$ parameter BPS black hole solutions with all different $Q_I$'s were discovered in \cite{Hristov:2019mqp}, whose physics is successfully described by our entropy function \eqref{S-AdS4}.

\section{AdS$_6$ black holes}

\subsection{Black hole solutions}

In this section, we study the supersymmetric AdS$_6$ black holes, and find an entropy function which accounts for their physics. We construct an entropy function for the solution of \cite{Chow:2008ip}. The solution may be regarded as describing BPS states of any large $N$ 5d SCFT dual. For instance, as our favorite example, results in this section may be understood
in the context of massive type IIA string theory on warped AdS$_6\times S^4/\mathbb{Z}_2$ product background.
This system is dual
to 5d $\mathcal{N}=1$ SCFT living on $N$ D4-branes probing the O8-D8 system \cite{Seiberg:1996bd}.
The 5d SCFT dual has a gauge theory description, with $Sp(N)$ gauge group, rank $2$ antisymmetric hypermultiplet, and $N_f\leq 7$ fundamental hypermultiplets. However, we expect that our general
analysis can be embedded to AdS$_6$ black holes in the backgrounds of
\cite{DHoker:2016ujz,Apruzzi:2014qva,Kim:2015hya}.

The 6d $\mathcal{N}=(1,0)$ $SU(2)$ gauged supergravity was obtained
by a consistent Kaluza-Klein truncation of massive type IIA supergravity on
$S^4/\mathbb{Z}_2$ \cite{Cvetic:1999un}. In \cite{Chow:2008ip}, the charged rotating AdS$_6$ black hole solution in this gauged supergravity was obtained. It has four kinds of conserved quantities: mass $E$, two angular momenta $J_1, J_2$, which describe the orthogonal 2-plane
rotations on $S^4$ in global AdS$_6$, and one $U(1) \subset SU(2)$ electric charge $Q$.
They are given in terms of four parameters $m, a, b, \delta$ of the solution as
\cite{Chow:2008ip}
\begin{equation}
\begin{aligned}
&E=\frac{2\pi m}{3G\Xi_a \Xi_b} \left[\frac{1}{\Xi_a}+\frac{1}{\Xi_b}+\sinh^2 \delta\left(1+\frac{\Xi_a}{\Xi_b}+\frac{\Xi_b}{\Xi_a}\right)  \right]\ ,\ \
Q=\frac{\pi m}{G\Xi_a \Xi_b}\sinh 2\delta\ ,\\
&J_1=\frac{2\pi ma}{3G\Xi_a^2 \Xi_b}(1+\Xi_b \sinh^2 \delta)\ , \ \
J_2=\frac{2\pi mb}{3G\Xi_a \Xi_b^2}(1+\Xi_a \sinh^2 \delta)\ ,
\end{aligned}
\end{equation}
where $\Xi_a=1-a^2g^2$ and $\Xi_b=1-b^2g^2$. The entropy is given by
\begin{equation}
S=\frac{2\pi^2\left[(r_+^2+a^2)(r_+^2+b^2)+2mr_+ \sinh^2 \delta\right]}{3G\Xi_a \Xi_b}\ .
\end{equation}
The event horizon is located at $r=r_+$. Here, $G$ is the 6d Newton constant. (In \cite{Chow:2008ip}, the unit $G=1$ is used.) $g$ is a gauge coupling constant in 6d gravity,
setting the inverse-radius of AdS$_6$.

This black hole solution admits the supersymmetric limit without naked closed timelike curves. The BPS condition
\begin{equation}\label{BPS-energy-AdS6}
E=gJ_1+gJ_2+Q
\end{equation}
is satisfied if
\begin{equation}\label{BPS-AdS6}
e^{2\delta}=1+\frac{2}{(a+b)g}\ .
\end{equation}
In addition, a smooth horizon exists only if
\begin{equation}\label{regular-AdS6}
m=\frac{(a+b)^2 (1+ag)(1+bg) (2+ag+bg)}{2(1+ag+bg)} \sqrt{\frac{ab}{1+ag+bg}}
\end{equation}
is satisfied. The horizon is located at
\begin{equation}
r_+=\sqrt{\frac{ab}{1+ag+bg}}\ .
\end{equation}
Taking the BPS limit, the entropy of the supersymmetric black hole is given by
\begin{equation}\label{BPS AdS6 black hole entropy}
S=\frac{2\pi^2 ab(a+b)}{3gG(1-ag)(1-bg)(1+ag+bg)}\ .
\end{equation}
The two conditions (\ref{BPS-AdS6}), (\ref{regular-AdS6}) leave two independent parameters among $m,a, b, \delta$. Even after restricting $E$ as (\ref{BPS-energy-AdS6}) from the BPS condition, the remaining charges $J_1,J_2,Q$ carried by the supersymmetric black holes will satisfy a
charge relation. Equivalently, together with $S$, we find the following two relations:
\begin{equation}\label{charge relation-AdS6 black hole}
\begin{aligned}
&S^3-\frac{2\pi^2}{3g^4G}S^2-12\pi^2\left(\frac{Q}{3g}\right)^2S+\frac{8\pi^4}{3g^4G}J_1 J_2=0\ , \\
&\frac{Q}{3g}S^2+\frac{2\pi^2}{9g^4G}(J_1+J_2)S-\frac{4\pi^2}{3}\left(\frac{Q}{3g}\right)^3=0\ .
\end{aligned}
\end{equation}
Since these equations determine $S$ twice, one will get a charge relation between $J_1, J_2, Q$ from the compatibility of two equations. Explicitly, one may take the unique positive solution of the second equation and insert it to the first equation, to get the charge relation.

The black hole chemical potentials and the free energy $F$ satisfy
\begin{equation}
S=-T^{-1}F+T^{-1}E-T^{-1}\Omega_1 J_1 -T^{-1}\Omega_2 J_2-T^{-1}\Phi Q\ ,
\end{equation}
where $T$ is the temperature, $\Omega_1, \Omega_2$ are the angular velocities, and $\Phi$ is the electrostatic potential. The temperature of the supersymmetric black hole is zero in the
BPS smooth horizon limit,
\begin{equation}
\hspace*{-.3cm}
T=\frac{2r_+^2(1\!+\!g^2r_+^2)(2r_+^2\!+\!a^2\!+\!b^2)\!-\!(1\!-\!g^2r_+^2)
(r_+^2+a^2)(r_+^2+b^2)\!+\!8mg^2r_+^3\sinh^2 \delta\!-\!4m^2g^2\sinh^4\delta}{4\pi r_+[(r_+^2+a^2)(r_+^2+b^2)+2mr_+\sinh^2 \delta]} \rightarrow 0\ .
\end{equation}
The other chemical potentials in the BPS limit are given by
\begin{equation}
\begin{aligned}
&\Omega_1=a\frac{(1+g^2r_+^2)(r_+^2+b^2)+2mg^2r_+\sinh^2\delta}{(r_+^2+a^2)(r_+^2+b^2)+2mr_+\sinh^2 \delta} \rightarrow g\ , \\
&\Omega_2=b\frac{(1+g^2r_+^2)(r_+^2+a^2)+2mg^2r_+\sinh^2\delta}{(r_+^2+a^2)(r_+^2+b^2)+2mr_+\sinh^2 \delta} \to g\ , \\
&\Phi=\frac{mr_+\sinh 2\delta}{(r_+^2+a^2)(r_+^2+b^2)+2mr_+\sinh^2 \delta} \to 1\ .
\end{aligned}
\end{equation}
Similar to the analysis in section \ref{AdS4-BH}, the following limits exist,
\begin{equation}
F_{\textrm{BPS}}=\lim_{T\to 0}(T^{-1}(F-\Delta E))\ , \; \omega_i=-\lim_{T\to 0}\left(T^{-1}(\Omega_i-g)\right)\ , \; \Delta=-\lim_{T\to 0}\left(T^{-1}(\Phi-1)\right)\ ,
\end{equation}
where $\Delta E\equiv E-Q-gJ_1-gJ_2$.
Then, in the zero temperature BPS limit, one obtains
\begin{equation}
S=-F_{\textrm{BPS}}+\omega_1 J_1 +\omega_2 J_2+\Delta Q\ .
\end{equation}
Using the computed expressions for $\omega_i,\Delta$, one finds that
\begin{equation}\label{chemical-relation-black-hole}
\omega_1+\omega_2= 3g \Delta\ .
\end{equation}
Again, this is the alternative statement of the charge relation of $J_1, J_2, Q$.

\subsection{Entropy function}

We now present an entropy function which encodes the physics of the BPS
black holes presented in the previous subsection. The entropy function is given by
\begin{equation}\label{AdS6 entropy function}
S=- i \frac{\pi}{81g^4 G} \frac{\Delta^3}{\omega_1 \omega_2}+\Delta Q + \omega_1 J_1 +\omega_2 J_2 + \lambda\Big( \Delta-\omega_1-\omega_2-2\pi i \Big)\ ,
\end{equation}
where $G$ is the 6d Newton constant as before.
Having in mind the concrete example of massive IIA supergravity on warped AdS$_6\times S^4/\mathbb{Z}_2$ background,
one would find $\frac{1}{g^4G} = \frac{27\sqrt{2}}{5\pi} \frac{N^{\frac{5}{2}}}{\sqrt{8-N_f}}$ \cite{Choi:2019miv}. In that case, a microscopic derivation of the entropy function (\ref{AdS6 entropy function}) from the CFT$_5$ dual was studied in \cite{Choi:2019miv}, in the Cardy limit $\omega_{1,2} \to 0$. 
Here, we introduced the Lagrange multiplier $\lambda$ to extremize $S$ in $\Delta, \omega_1, \omega_2$ subject to the constraint
\begin{equation}\label{chemical-relation-entropy-function}
\Delta-\omega_1-\omega_2=2\pi i\ .
\end{equation}
Differentiating with respect to the chemical potentials, one obtains
\begin{equation}\label{AdS6-extremum}
  \lambda+Q= i \frac{\pi}{27g^4G} \frac{\Delta^2}{\omega_1 \omega_2}\ \ ,\ \ \
  \lambda-J_1= i \frac{\pi}{81g^4G} \frac{\Delta^3}{\omega_1^2 \omega_2}\ \ ,\ \ \
  \lambda-J_2= i \frac{\pi}{81g^4G} \frac{\Delta^3}{\omega_1 \omega_2^2}\ .
\end{equation}
Inserting these back to the original entropy function formula, one obtains
\begin{equation}
S=-2\pi i \lambda.
\end{equation}
Multiplying the last two equations of (\ref{AdS6-extremum}), one obtains
\begin{equation}
(\lambda-J_1)(\lambda-J_2)=-\left(\frac{\pi}{81g^4G}\right)^2 \frac{\Delta^6}{\omega_1^3 \omega_2^3}=- i \frac{3g^4G}{\pi}(\lambda+Q)^3\ .
\end{equation}
Hence, one obtains
\begin{equation}\label{S-AdS6}
\left(\frac{S}{2\pi i}+J_1\right)\left(\frac{S}{2\pi i}+J_2\right)= i \frac{3g^4G}{\pi}\left(\frac{S}{2\pi i}-Q\right)^3\ .
\end{equation}

As in our section \ref{AdS4} and \cite{Choi:2018hmj}, we dismiss
${\rm Im}(S)$, focussing on ${\rm Re}(S)$ as our entropy. However, again note that
all known supersymmetric AdS$_6$ black holes have a charge relation. This charge relation
will coincide with the condition ${\rm Im}(S)=0$ at the saddle point.
So we demand real $S$ for real charges $Q, J_1, J_2$. Then, (\ref{S-AdS6})
is separated into two real equations as follows:
\begin{eqnarray}\label{charge relation-entropy function}
\begin{aligned}
&S^3-\frac{2\pi^2}{3g^4G}S^2-12\pi^2Q^2S+\frac{8\pi^4}{3g^4G}J_1J_2=0\ , \\
&QS^2+\frac{2\pi^2}{9g^4G}(J_1+J_2)S-\frac{4\pi^2}{3}Q^3=0\ .
\end{aligned}
\end{eqnarray}
These equations determine $S$ twice as functions of charges. Therefore, from the compatibility of two equations, one obtains a charge relation of $Q, J_1, J_2$. These two equations of $S, Q, J_1, J_2$ (\ref{charge relation-entropy function}), derived from the entropy function (\ref{AdS6 entropy function}), are exactly the same as those from the supersymmetric black holes (\ref{charge relation-AdS6 black hole}). Note that the charges and chemical potentials of the entropy function (\ref{AdS6 entropy function}) are related to those of the black holes as
\begin{equation}
\begin{aligned}
&S_{\textrm{BH}}=S, \; J_{i, \textrm{BH}}=J_{i}, \; \frac{1}{3g}Q_{\textrm{BH}}=Q\ , \\
&\omega_{i, \textrm{BH}}=\textrm{Re}(\omega_{i}), \; 3g \, \Delta_{\textrm{BH}}=\textrm{Re}(\Delta)\ .
\end{aligned}
\end{equation}
The subscripts `BH' denote the black hole quantities, while the others are the quantities used in the entropy function. One can also realize that the relation between the chemical potentials in the entropy function (\ref{chemical-relation-entropy-function}) is equivalent to that of the supersymmetric black holes (\ref{chemical-relation-black-hole}).

Thus, our entropy function (\ref{AdS6 entropy function}) indeed reproduces the Bekenstein-Hawking entropy of the supersymmetric AdS$_6$ black holes (\ref{BPS AdS6 black hole entropy}),
and also their chemical potentials.

\section{Concluding remarks}

In this note, we presented the entropy functions of supersymmetric black holes in
AdS$_4$ and AdS$_6$. Complicated black hole quantities can be very concisely understood
from simple extremization principles of these entropy functions. Considering the
inspirations given by the similar
entropy functions in AdS$_5$ \cite{Hosseini:2017mds} or AdS$_7$
\cite{Hosseini:2018dob} to their microscopic studies, we expect that our entropy functions
will also play similar roles.

The entropy function in AdS$_4$, presented in section \ref{AdS4}, was actually motivated by
our microscopic study \cite{Choi:2019zpz} of BPS states in the M2-brane QFT (see, e.g. \cite{Bagger:2006sk,Gustavsson:2007vu,Aharony:2008ug})
from its index \cite{Kim:2009wb}. In particular, \cite{Choi:2019zpz} derived the entropy function (\ref{entropy-ftn-AdS4}) for large AdS$_4$ black holes, from the index of the radially quantized SCFT on M2-branes, where the condensation of the magnetic monopole operators gives rise to the novel deconfined $N^{\frac{3}{2}}$ degrees of freedom.

One can also derive our results on AdS$_6$ black holes from 5d SCFT duals, for instance, the strong coupling limits of 5d gauge theories on the D4-D8-O8 system \cite{Choi:2019miv}.
The indices of such SCFTs on $S^4\times\mathbb{R}$ were explored in \cite{Kim:2012gu},
where the problem reduced to studies of the 5d instanton partition functions: see, e.g.
\cite{Nekrasov:2002qd,Nekrasov:2004vw,Hwang:2014uwa}.
Later, \cite{Choi:2019miv} derived the entropy function (\ref{AdS6 entropy function}) for large AdS$_6$ black holes, from the indices of such 5d SCFTs and their orbifold theories, where the instanton solitons play subtle roles to realize deconfined $N^{\frac{5}{2}}$ degrees of freedom. Furthermore, while AdS$_6$ black hole solution known to date has only one electric charge dual to R-charge, \cite{Choi:2019miv} obtained a more general form of the entropy function, which describes AdS$_6$ black holes carrying various electric charges, dual to R-charge, mesonic charge and baryonic charges, yet to be discovered. For example, when the black hole has one more electric charge dual to the mesonic charge, the numerator $\sim \Delta^3$ of our entropy function (\ref{AdS6 entropy function}) is refined to $[(\Delta+\hat{m})(\Delta-\hat{m})]^{\frac{3}{2}}$, where $m\equiv \hat{m}+2\pi i$ is the chemical potential conjugate to the mesonic charge.

One may think of generalizations of our results on AdS$_4\times S^7$, to more general
4d $\mathcal{N}=2$ gauged supergravity models arising from string or M-theory.
To see a natural possibility of generalization, note that the numerator $\sim\sqrt{\Delta_1\Delta_2\Delta_3\Delta_4}$ of our entropy function
(\ref{entropy-ftn-AdS4}) is the homogeneous degree $2$ prepotential of the
$U(1)^4$ supergravity \cite{Cvetic:1999xp}. The prepotential is the square root
of a degree $4$ polynomial. See, e.g.
\cite{Gauntlett:2009zw,Lee:2014rca} for such structures in
other backgrounds. We conjecture that, for BPS black holes in 4d
$\mathcal{N}=2$ gauged supergravity, an entropy function like (\ref{entropy-ftn-AdS4})
can be constructed by replacing the numerator by the prepotential of the theory.
Recently, such an entropy function was found in \cite{Hristov:2019mqp}, and also microscopically studied in \cite{Choi:2019dfu} from the indices of CFT$_3$ duals in the Cardy limit $\omega \to 0$.

Here, note that similar prepotentials appeared in the entropy functions of 
magnetic/dyonic AdS$_4$ black holes \cite{Benini:2015eyy} (see also \cite{Hosseini:2016tor}).
The unifying underlying structures for all these entropy functions were microscopically studied in \cite{Choi:2019dfu} from CFT$_3$ duals. We finally note that the entropy functions of 
electric AdS$_6$ black holes in our paper also appear to have some similarities with 
magnetized black holes in AdS$_6$, with boundaries replaced by more general 4-manifolds
\cite{Hosseini:2018uzp,Suh:2018tul}.

\vskip 0.5cm

\hspace*{-0.8cm} {\bf\large Acknowledgements}
\vskip 0.2cm

\hspace*{-0.75cm} We thank Dongmin Gang,
Hee-Cheol Kim, Joonho Kim, Kimyeong Lee and Jaemo Park
for helpful discussions. This work is supported in part
by the National Research Foundation of Korea (NRF) Grant 2018R1A2B6004914 (SC, SK, JN),
NRF-2017-Global Ph.D. Fellowship Program (SC), the ERC-STG grant 637844-HBQFTNCER (CH), the INFN (CH), 
and Hyundai Motor Chung Mong-Koo Foundation (JN).

%\pagebreak

%\appendix

%\pagebreak

\end{document}